\documentclass{Interspeech2024}

\interspeechcameraready 
\usepackage[table,xcdraw]{xcolor}
\usepackage{colortbl}
\definecolor{Gray}{gray}{0.9}
\usepackage{cite}           % compressed citation lists
\usepackage{amsmath}
\usepackage{amssymb}% http://ctan.org/pkg/amssymb
\usepackage{pifont}% http://ctan.org/pkg/pifont
\newcommand{\cmark}{\ding{51}}%
\newcommand{\xmark}{\ding{55}}%
\usepackage{multirow}

\title{Whisper-Flamingo: Integrating Visual Features into Whisper\\ for Audio-Visual Speech Recognition and Translation}

\name[affiliation={1}]{Andrew}{Rouditchenko}
\name[affiliation={1}]{Yuan}{Gong}
\name[affiliation={2,3}]{Samuel}{Thomas}
\name[affiliation={2,3}]{Leonid}{Karlinsky}
\name[affiliation={3,4}]{Hilde}{Kuehne}
\name[affiliation={2,3}]{Rogerio}{Feris}
\name[affiliation={1}]{James}{Glass}

\address{$^1$MIT, USA \hspace{2mm}
  $^2$IBM Research AI, USA \hspace{2mm}
  $^3$MIT-IBM Watson AI Lab, USA \hspace{2mm} \\
  $^4$University of Bonn, Germany \hspace{2mm}}
\email{roudi@mit.edu}

\keywords{audio-visual speech recognition, noise-robust}

\begin{document}

\maketitle

% the abstract here must exactly match the abstract entered into the paper submission system
\begin{abstract}
    
    % 1000 characters. ASCII characters only. No citations.
% v4, incorporating feedback from Yuan
Audio-Visual Speech Recognition (AVSR) uses lip-based video to improve performance in noise. 
Since videos are harder to obtain than audio, the video training data of AVSR models is usually limited to a few thousand hours. 
In contrast, speech models such as Whisper are trained with hundreds of thousands of hours of data, and thus learn a better speech-to-text decoder. 
The huge training data difference motivates us to adapt Whisper to handle video inputs. 
Inspired by Flamingo which injects visual features into language models, we propose Whisper-Flamingo which integrates visual features into the Whisper speech recognition and translation model with gated cross attention. 
Our models achieve state-of-the-art ASR WER (0.68\%) and AVSR WER (0.76\%) on LRS3, and state-of-the-art ASR WER (1.3\%) and AVSR WER (1.4\%) on LRS2.
Audio-visual Whisper-Flamingo outperforms audio-only Whisper on English speech recognition and En-X translation for 6 languages in noisy conditions.
Moreover, Whisper-Flamingo is versatile and conducts all of these tasks using one set of parameters, while prior methods are trained separately on each language.

\end{abstract}

\section{Introduction}

In recent years, major improvements in Automatic Speech Recognition (ASR) performance have been achieved by models trained on large-scale data~\cite{radford2023robust,zhang2023google}, but performance still declines in noise~\cite{gong23d_interspeech}.
To enhance performance in noise, Audio-Visual Speech Recognition (AVSR) uses lip-based video in addition to audio inputs.
Self-Supervised Learning (SSL) methods such as AV-HuBERT~\cite{shi2022learning} pre-train on large-scale datasets of unlabeled videos and fine-tune on a few hundred hours of labeled videos to perform noise-robust AVSR.
However, due to the difficulty in collecting publicly accessible videos, these models are usually trained on only a few thousand hours of data.

To overcome the lack of video data, recent methods fine-tune audio-only models pre-trained on hundreds of thousands of hours of audio for audio-visual speech recognition~\cite{pan-etal-2022-leveraging,may2023audio,simic2023self}.
The results show that such audio models combined with audio-visual fine-tuning on a few hundred hours of videos can approach the performance of video models pre-trained on thousands of hours of video~\cite{may2023audio}.
However, these methods often train video models and text decoders from scratch on only a few hundred hours of data, which is suboptimal compared to training on large-scale data.
Furthermore, only English data has been used.

In this work, we propose to integrate visual features from AV-HuBERT into Whisper~\cite{radford2023robust}, an audio-only model trained on 680k hours of speech with a strong \textit{multilingual} decoder.
Compared to the prior audio-visual adaptation methods, our video model and text decoder are pre-trained with large-scale data.
This allows our method to perform well on audio-visual speech translation, a task not explored by prior methods~\cite{pan-etal-2022-leveraging,may2023audio,simic2023self}.

How to fuse modalities effectively in multi-modal models is an ongoing research question.
One recent work, Flamingo~\cite{alayrac2022flamingo}, fuses visual features into text-only language models using gated cross attention and fine-tuning on a paired text-image dataset.
The gated cross attention layers are initialized to the identity function and learn to attend to the visual features during fine-tuning.
These layers have been shown to generalize to different modality pairs; Audio Flamingo~\cite{kong2024audio} recently applied them for text-audio reasoning.
Inspired by this method, we propose Whisper-Flamingo which inserts gated cross attention layers into Whisper's decoder and enables Whisper to use lip-based features for speech recognition.

On the English (En) LRS3 video dataset~\cite{afouras2018lrs3}, our models achieve State-of-the-Art (SOTA) ASR WER (0.68\%) and AVSR WER (0.76\%).
On LRS2~\cite{afouras2018adeep}, our models achieve SOTA ASR WER (1.3\%) and AVSR WER (1.4\%).
Our novel audio-visual Whisper-Flamingo significantly outperforms the audio-only Whisper baseline in noise.
Moreover, Whisper-Flamingo achieves competitive noise-robust results compared to prior audio-visual models.
Next, we demonstrate Whisper's multilingual capabilities by extending Whisper-Flamingo for En-X translation on the MuAViC dataset~\cite{anwar23_interspeech}.
Our model performs En transcription and En-X translation into 6 other languages, while the previous audio-visual SOTA requires fine-tuning on each language separately.
Once again, Whisper-Flamingo significantly outperforms audio-only Whisper in noise for both En transcription and En-X translation.
Code and models at \break \begin{footnotesize}  \url{https://github.com/roudimit/whisper-flamingo} \end{footnotesize}

\begin{figure*}[t]
    \centering
    \includegraphics[width=0.95\textwidth]{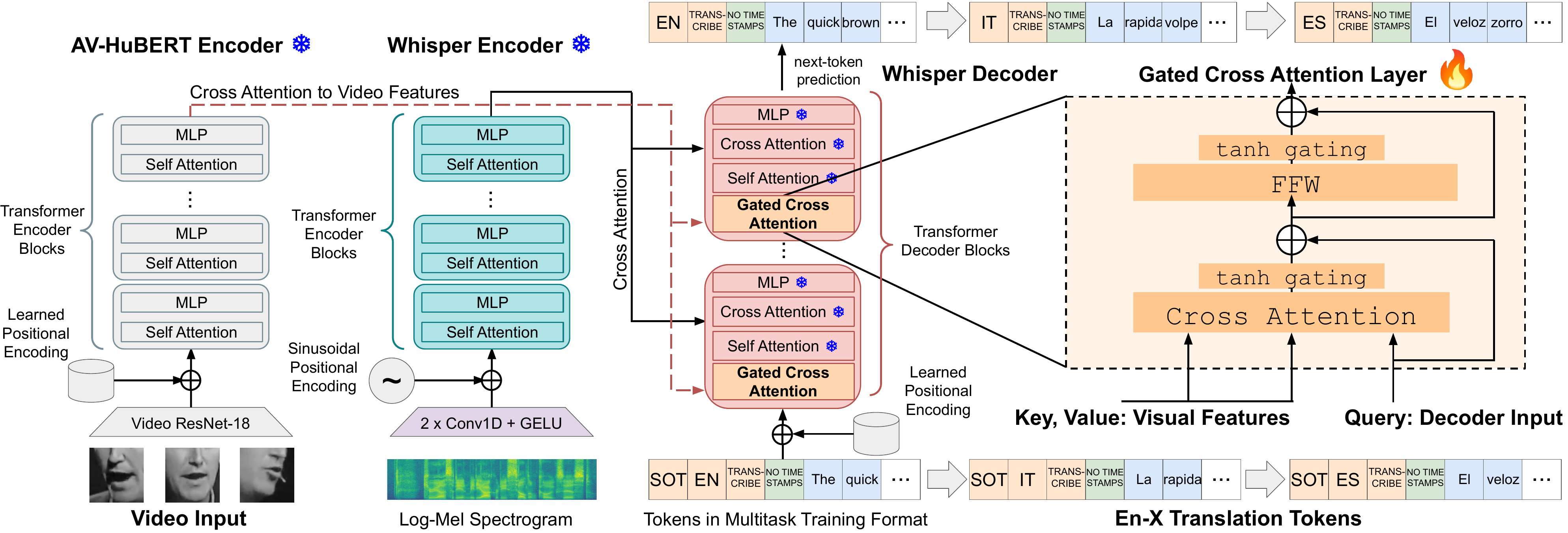}
    \caption{
    Diagram of Whisper-Flamingo based on Whisper~\cite{radford2023robust} and Flamingo~\cite{alayrac2022flamingo}. 
    We first fine-tune all of Whisper's parameters using English audio for English transcription and En-X translation.
    To train Whisper-Flamingo, we freeze the audio model, add gated cross attention layers into Whisper's decoder attending to visual features from AV-HuBERT, and train the model on audio-visual inputs.
    }
    \label{fig:model_fig}
    \vspace{-0.4cm}
\end{figure*}

\section{Method}
In this section, we review audio-visual fusion methods for AVSR, and then explain our method.
Two common fusion methods are early and late fusion.
In early fusion, both modalities are first separately processed by light-weight encoders and then combined with feature addition or concatenation and used as input to an audio-visual Transformer~\cite{vaswani2017attention,afouras2018adeep}.
Both SSL models~\cite{shi2022learning,hsu2022u,zhang2023self} and fully-supervised models~\cite{makino2019recurrent,serdyuk22_interspeech,rouditchenko2023av} use this design.
In late fusion, audio and video are processed separately by Transformer encoders, and afterwards features are fused with an MLP.
The audio-visual features are then passed to a linear-layer or Transformer decoder.
This approach is common for fully-supervised models~\cite{petridis2018end,ma2021end,ma2023auto}.
Both early and late fusion need identical audio and visual feature rates so that they can be fused at each time step; a common design is to match the video's frame rate by downsampling the audio features to 25 Hz.

\begin{table}[t]
    \centering
    \caption{Hyperparameter summary. 
    We first train Whisper-Large FT (Fine-tune) with audio-only, then use it to initialize Whisper-Flamingo.
    A=audio, AV=audio-visual. \textdagger=per sample.}
    \label{tab:hyperparameters}
    \vspace{-3mm}
\resizebox{0.97\linewidth}{!}{%

\begin{tabular}{lcccc}
\toprule
 & Whisper- & Whisper- & Whisper- & Whisper- \\
 & Large FT & Large FT & Flamingo & Flamingo \\
\midrule
Test Modalities & A & A & AV & AV \\
En Recognition & \cmark & \cmark & \cmark & \cmark \\
En-X Translation & \xmark & \cmark & \xmark & \cmark \\
GPUs & 1 & 4 & 1 & 4 \\
Total Params. & 1.55B & 1.55B & 2.5B & 2.5B \\
AV-HuBERT Params. & - & - & 325M & 325M \\
Gated X-Attn Params. & - & - & 630M & 630M \\
Trainable Params. & 1.55B & 1.55B & 631M & 631M \\
Warmup Steps & 1k & 1k & 5k & 5k \\
Total Steps & 90k & 225k & 20k & 40k \\
Learning Rate & $5\times10^{-6}$ & $5\times10^{-6}$ & $1\times10^{-4}$ & $1\times10^{-4}$ \\
Batch per GPU (s) & 80 & 40 & 160 & 30 \\
Max Length (s) \textdagger & 10 & 10 & 15 & 15 \\
Max Characters \textdagger & 350 & 300 & 350 & 250 \\
\bottomrule
\end{tabular}%
}
\vspace{-0.4cm}
\end{table}
Most methods which adapt pre-trained audio-only models for AVSR through audio-visual fine-tuning use early fusion.
FAVA~\cite{may2023audio} adapts BEST-RQ~\cite{chiu2022self}, an audio self-supervised model, through early fusion with a video model trained from scratch.
Adaptive AV~\cite{simic2023self} prepends Whisper with an audio-visual Transformer to output a de-noised spectrogram, but does not use visual features directly in Whisper.
However, we found gated cross attention with features from pre-trained AV-HuBERT worked better than early fusion.
% Note that separate research focuses on using visual features from images or instructional videos for AVSR, where the visuals provide context and are only loosely synchronized with the audio~\cite{sun2016look,caglayan2019multimodal,ghorbani2021listen,seo2023avformer,peng23d_interspeech}.
Note that separate research focuses on using visual features from images or instructional videos for AVSR, where the visuals provide context and are only loosely synchronized with the audio~\cite{seo2023avformer,peng23d_interspeech}.

We propose to adapt Whisper's decoder with visual features from AV-HuBERT using gated cross attention, 
as shown in Figure~\ref{fig:model_fig}.
Each of Whisper's decoder blocks consists of a self-attention layer, cross attention layer attending to the audio features, and a Multi-Layer Perceptron (MLP).
Based on Flamingo~\cite{alayrac2022flamingo}, the gated cross attention layer is defined as follows, where $\mathbf{x}$ is the input to the decoder block, $\mathbf{v}$ are the visual features, $\mathrm{Attn}$ is multi-head cross attention, $\mathrm{LN}$ is Layernorm~\cite{ba2016layer}, and $\mathrm{FFW}$ is an MLP:
\begin{align}
    \mathbf{x^\prime}     &= \mathbf{x} + \mathrm{tanh}(\mathrm{\alpha_{xattn}}) \times \mathrm{Attn}(\mathrm{LN}(\mathbf{x}), \mathbf{v})  \\
    \mathbf{y} &= \mathbf{x^\prime} + \mathrm{tanh}(\mathrm{\alpha_{mlp}}) \times \mathrm{FFW}(\mathrm{LN}(\mathbf{x^\prime})) 
\end{align}
The learnable parameters $\mathrm{\alpha_{xattn}}$ and $\mathrm{\alpha_{mlp}}$ are initialized to 0 so that the layers initially function as the identity since $\tanh(0) = 0$.
Through audio-visual fine-tuning, the model adjusts the weights of $\mathrm{\alpha_{xattn}}$ and $\mathrm{\alpha_{mlp}}$ and learns to attend to the visual features.
We insert the gated cross attention layers in Whisper's decoder in the beginning of each block, before the self-attention layer. 
We tried to insert them in other orders within the decoder blocks, but the performance was slightly worse.
The full analysis is shown in the Appendix, Section~\ref{sec:appendix-x-attn}.
Note that since the gated cross attention separately attends to the video features, the audio and video features can have different feature rates (for example, 50 Hz and 25 Hz).

% V2 - beam search results
\begin{table}[t]
    \centering
    \caption{Fusion ablation with Whisper-Medium on LRS3. 
    We report results on the original test set (Clean) and with babble noise injected at 0-SNR (Noisy).
    A=audio, AV=audio-visual.
    }
    \label{tab:fusion}
    \vspace{-3mm}
\resizebox{0.77\linewidth}{!}{%
\begin{tabular}{lcrr}
\toprule
 & Test & \multicolumn{1}{c}{Clean} & \multicolumn{1}{c}{Noisy} \\
Model & Modalities & \multicolumn{1}{c}{WER$\downarrow$} & \multicolumn{1}{c}{WER$\downarrow$} \\
\midrule
Whisper, Zero-shot & A & 2.3 & 22.2 \\
Whisper, Fine-tuned & A & \textbf{1.9} & \textbf{12.6} \\
\midrule
Whisper-Early-Fusion & AV & 1.7 & 10.0 \\
Whisper-Late-Fusion & AV & 2.1 & 16.5 \\
Whisper-Flamingo & AV & \textbf{1.5} & \textbf{7.0} \\
\bottomrule
\end{tabular}%
}
\vspace{-0.4cm}
\end{table}

% V2 - only beam search results
\begin{table*}[t]
    \centering
    \caption{Results for English transcription on LRS3. 
    We report results on the original test set (Clean) and with babble noise added at 0-SNR (Noisy).
    Hours of unlabeled \& labeled audio-visual data used to train each model are shown.
    433h=LRS3, 1,759h=LRS3+VoxCeleb2.
    % A= Audio, AV= Audio-Visual.
    Noise dataset= dataset used to make babble noise for testing.
    Noisy WER from methods using different noise datasets are not directly comparable.
    Note\textsuperscript{\textdagger} that u-HuBERT~\cite{hsu2022u}, AV-HuBERT~\cite{shi22_interspeech}, and CMA~\cite{kim2024learning} use a different noise file than us.
    *=noisy results were not reported using 0-SNR.
    }
    \label{tab:lrs3}
    \vspace{-3mm}
\resizebox{0.95\linewidth}{!}{%
\begin{tabular}{lcrrrrrr}
\toprule
 & Noise & \multicolumn{2}{c}{AV Training Hrs} & \multicolumn{2}{c}{ASR WER$\downarrow$} & \multicolumn{2}{c}{AVSR WER$\downarrow$} \\
Model & \cellcolor[HTML]{FFFFFF}Dataset & Unlabeled & Labeled & Clean & Noisy & Clean & Noisy \\
\hline
\rowcolor{Gray} \multicolumn{8}{c}{\textit{Audio-Visual methods trained only on transcribed videos}} \\
VIT 3D~\cite{serdyuk22_interspeech}  & NoiseX & - & 90k & 1.6 & \textbf{6.1} & 1.6 & 2.9 \\
LP Conformer~\cite{chang2022on} & NoiseX & - & 100k & - & - & 0.9 & \textbf{1.9} \\
AutoAVSR~\cite{ma2023auto}  & NoiseX & - & 1,902 / 3,448 & 1.0 / 1.0 & * & 1.0 / 0.9 & * \\
Fast Conformer~\cite{burchi2024multilingual}  & NoiseX & -  & 435 / 3,116 & 1.6 / \textbf{0.7} & * & 1.7 / \textbf{0.8} & * \\
\hline
\rowcolor{Gray} \multicolumn{8}{c}{\textit{Audio-Visual SSL methods}} \\
AV2vec~\cite{zhang2023self}  & MUSAN & 433 & 433 & 2.7 & 19.5 & 2.5 & 6.7 \\
AV-BEST-RQ~\cite{may2023audio}  & NoiseX & 1,759 & 433 & - & - & 2.1 & 6.8 \\
% BRAVEn~\cite{haliassos2024braven}  & NoiseX & 2,649 & 433 & \textbf{1.1} & 18.2 & - & 15.0 \\
BRAVEn~\cite{haliassos2024braven}  & NoiseX & 2,649 & 433 & \textbf{1.1} & - & - & - \\
AV-HuBERT~\cite{shi22_interspeech}  & LRS3{\textdagger} & 1,759 & 433 & 1.6 & \textbf{15.7} & 1.4 & 5.8 \\
u-HuBERT~\cite{hsu2022u}  & LRS3{\textdagger} & 1,759 & 433 & 1.4 & 19.3 & \textbf{1.3} & 4.6 \\
CMA~\cite{kim2024learning} & LRS3{\textdagger} & 1,759 & 433 & - & - & 1.5 & \textbf{4.4} \\
\hline
\rowcolor{Gray} \multicolumn{8}{c}{\textit{Audio-Only pre-train + audio-visual fine-tune methods}} \\
Adaptive AV~\cite{simic2023self}  & MUSAN & 400 & 30 & - & - & 2.3 & 16.3 \\
FAVA-USM~\cite{may2023audio}  & NoiseX & - & 433 & - & - & 1.3 & 6.2 \\
Llama-AVSR~\cite{cappellazzo2024large}  & NoiseX & 1,759 & 433 / 1,759 & 1.1 / \textbf{0.81} & \textbf{12.3} / - & 0.95 / \textbf{0.77} & \textbf{4.2} / - \\
\hline
\rowcolor{Gray} \multicolumn{8}{c}{\textit{Our audio-only Whisper Zero-shot baseline}} \\
Whisper-Large, Zero-shot, Beam 1  & LRS3 & - & - & 2.3 & 23.3 & - & - \\
$\hookrightarrow$ w/ Beam search 15  & LRS3 & - & - & \textbf{2.1} & \textbf{20.8} & - & - \\
\hline
\rowcolor{Gray} \multicolumn{8}{c}{\textit{Our models trained \textbf{without} noise}} \\
Whisper-Large, Fine-tuned w/o noise, Beam 1  & LRS3 & - & 433 / 1,759 & 1.0 / \textbf{0.68} & \textbf{21.6} / 23.6 & - & - \\
$\hookrightarrow$ w/ Beam search 15  & LRS3 & - & 433 / 1,759 & 1.3 / 1.5 & 23.1 / 24.1 & - & - \\
Whisper-Flamingo, Fine-tuned w/o noise, Beam 1  & LRS3 & 1,759 & 433 / 1,759 & - & - & 1.1 / \textbf{0.76} & 13 / 14.9 \\
$\hookrightarrow$ w/ Beam search 15  & LRS3 & 1,759 & 433 / 1,759 & - & - & 1.1 / 1.8 & \textbf{12.8} / 15.9 \\
\hline
\rowcolor{Gray} \multicolumn{8}{c}{\textit{Our models trained \textbf{with} noise}} \\
Whisper-Large, Fine-tuned w/ noise, Beam 1  & LRS3 & - & 433 / 1,759 & 1.1 / \textbf{0.85} & 12.4 / 11.9 & - & - \\
$\hookrightarrow$ w/ Beam search 15  & LRS3 & - & 433 / 1,759 & 2.3 / 2.0 & 11.7 / \textbf{11.1} & - & - \\
Whisper-Flamingo, Fine-tuned w/ noise, Beam 1  & LRS3 & 1,759 & 433 / 1,759 & - & - & 1.0 / \textbf{0.86} & 7.2 / 6.1 \\
$\hookrightarrow$ w/ Beam search 15  & LRS3 & 1,759 & 433 / 1,759 & - & - & 1.5 / 2.0 & \textbf{5.6} / \textbf{5.6} \\
\bottomrule
\end{tabular}%
}
\vspace{-0.4cm}
\end{table*}

\noindent \textbf{Training pipeline.}
Before adding gated cross attention, we first fine-tune all layers of the audio-only Whisper model to adapt it to the domain of interest (denoted as Whisper Fine-tuned).
We also add noise during fine-tuning to increase the noise-robustness.
We use the standard cross-entropy loss between the model's predicted transcript and the ground-truth tokens.
To train Whisper-Flamingo, we \textit{freeze} the fine-tuned Whisper, insert the gated cross attention layers, and fine-tune the model with audio-visual inputs.
The gated cross attention layers and a linear layer on top of the visual features are trained from scratch, while all other parameters are frozen.
The new layers can therefore be seen as a (large) set of adaptors~\cite{houlsby2019parameter}: removing them results in the audio-only Whisper weights.

\noindent \textbf{From English to Multilingual.}
Whisper was trained for multilingual transcription and X-En translation (multilingual audio to En text).
We tried Whisper-Flamingo on multilingual speech recognition and X-En translation using the videos in the MuAViC dataset~\cite{anwar23_interspeech} but found several issues.
Most languages in the dataset have less than a third of the hours of English data available, which makes training new layers from scratch difficult.
Also, the multilingual videos are longer on average than the English videos.
This causes increased GPU memory pressure and requires a reduced batch size, which also makes training difficult.
Therefore we focused on \textit{En-X} translation (English audio to multilingual text) and propose to handle multilingual recognition and translation in future work~\cite{cheng2023mixspeech,hong-etal-2023-intuitive,li2023parameter,burchi2024multilingual,han2024xlavs}.

Prior research shows that Whisper can be prompted for En-X translation, but it requires language-specific logit filtering and the performance can still be unsatisfactory~\cite{peng23d_interspeech}.
Since fine-tuning Whisper has been shown to enable transcription of unseen languages~\cite{rouditchenko23_interspeech}, we propose to fine-tune Whisper for En-X translation.
We fine-tune the audio model in a multi-task style to transcribe English audio and translate it to the other languages.
To train Whisper-Flamingo, we freeze the fine-tuned audio model, add the gated cross attention layers and the linear layer on top of the visual features, and train the model on audio-visual inputs.
~\section{Experiments on LRS3}
% v4 - only beam search
\begin{table*}[t]
    \centering
    \setlength{\tabcolsep}{8pt}
    \caption{Results for English transcription on LRS3 and En-X Translation on MuAViC.
    Babble noise is added at 0-SNR (Noisy).
    One Model= the model translates to all languages with one set of parameters.
    Test Mod.= inference modalities (Text: T, audio: A, video: V audio-visual: AV).
    Note\textsuperscript{\textdagger} that Bilingual AV-HuBERT~\cite{anwar23_interspeech} use a different noise file than us that was not publicly available so the results in noisy conditions are not directly comparable.
    }
    \label{tab:en-x}
    \vspace{-3mm}
\resizebox{0.90\linewidth}{!}{%
\begin{tabular}{lcccrrrrrrrr}
\toprule
 & Test & One & Noise & \multicolumn{1}{c}{En} & \multicolumn{1}{c}{El} & \multicolumn{1}{c}{Es} & \multicolumn{1}{c}{Fr} & \multicolumn{1}{c}{It} & \multicolumn{1}{c}{Pt} & \multicolumn{1}{c}{Ru} & \multicolumn{1}{c}{Avg} \\
\cmidrule{6-11}
Model & Mod. & Model & Dataset & \multicolumn{1}{c}{WER$\downarrow$} & \multicolumn{6}{c}{BLEU$\uparrow$} & \multicolumn{1}{c}{w/o En} \\
\hline
\rowcolor{Gray}\multicolumn{12}{c}{\textit{Text-to-Text Translation}} \\
Bilingual Transformer~\cite{anwar23_interspeech} & T & \xmark & - & - & \textbf{25.8} & \textbf{29.5} & \textbf{27.0} & \textbf{22.6} & \textbf{23.9} & \textbf{17.2} & \textbf{24.3} \\
M2M-100~\cite{fan2021beyond,anwar23_interspeech} & T & \cmark & - & - & 24.5 & 28.7 & 25.6 & 21.8 & 22.2 & 15.8 & 23.1 \\
\hline
\rowcolor{Gray}\multicolumn{12}{c}{\textit{Speech-to-Text Translation (Clean Audio)}} \\
Bilingual AV-HuBERT~\cite{anwar23_interspeech} & A & \xmark & - & - & \underline{23.0} & \underline{27.5} & 25.1 & 20.7 & 20.1 & 14.7 & 21.9 \\
Whisper-Small, Fine-tuned & A & \cmark & - & \underline{2.0} & 22.4 & 27.1 & 24.9 & 20.9 & \textbf{21.6} & \underline{15.6} & 22.1 \\
Whisper-Medium, Fine-tuned & A & \cmark & - & 2.1 & 22.9 & \underline{27.5} & \textbf{26.1} & \textbf{21.9} & \underline{21.4} & 15.1 & \underline{22.5} \\
Whisper-Large, Fine-tuned & A & \cmark & - & \textbf{1.5} & \textbf{23.7} & \textbf{27.9} & \underline{26.0} & \underline{21.8} & \underline{21.4} & \textbf{15.7 }& \textbf{22.7} \\
\midrule
Bilingual AV-HuBERT~\cite{anwar23_interspeech} & AV & \xmark & - & - & 23.4 & 26.6 & 25.3 & 20.7 & 20.5 & 14.6 & 21.9 \\
(\textbf{Ours}) Whisper-Flamingo (Small)  & AV & \cmark  & - & 2.0 & 22.6 & 27.0 & 24.7 & 20.7 & \underline{21.3} & 15.5 & 22.0 \\
(\textbf{Ours}) Whisper-Flamingo (Medium)  & AV & \cmark  & - & \underline{1.6} & \underline{23.8} & \textbf{28.0} & \textbf{26.1} & \textbf{22.5} & \underline{21.3} & \textbf{16.0} & \textbf{23.0} \\
(\textbf{Ours}) Whisper-Flamingo (Large)  & AV & \cmark  & - & \textbf{1.3} & \textbf{24.4} & \underline{27.9} & \underline{25.9} & \underline{22.1} & \textbf{21.8 }& \underline{15.7} & \underline{22.9} \\
\hline
\rowcolor{Gray}\multicolumn{12}{c}{\textit{Speech-to-Text Translation (Noisy Audio from MuAViC)}} \\
Bilingual AV-HuBERT~\cite{anwar23_interspeech} & A & \xmark & MuAViC\textdagger & - & 15.9 & 19.2 & 17.1 & 12.9 & 14.4 & 10.3 & 15.0 \\
Whisper-Small, Fine-tuned & A & \cmark & MuAViC & 17.3 & 17.5 & 20.1 & 19.4 & 15.3 & 16.3 & 11.8 & 16.7 \\
Whisper-Medium, Fine-tuned & A & \cmark & MuAViC & \underline{14.8} & \underline{18.1} & \underline{22.1} & \underline{19.8} & \underline{16.2} & \underline{17.3} & \underline{12.1} & \underline{17.6} \\
Whisper-Large, Fine-tuned & A & \cmark & MuAViC & \textbf{13.8} & \textbf{19.7} & \textbf{23.4} & \textbf{20.4} & \textbf{17.4} & \textbf{17.7} & \textbf{13.3} & \textbf{18.6} \\
\midrule
Bilingual AV-HuBERT~\cite{anwar23_interspeech} & AV & \xmark & MuAViC\textdagger & - & \textbf{22.7} & \underline{24.8} & \textbf{23.8} & \textbf{20.0} & \textbf{20.0} & \underline{13.7} & \textbf{20.8} \\
(\textbf{Ours}) Whisper-Flamingo (Small)  & AV & \cmark  & MuAViC & 10.7 & 19.0 & 22.1 & 21.1 & 17.1 & 18.3 & 13.2 & 18.5 \\
(\textbf{Ours}) Whisper-Flamingo (Medium)  & AV & \cmark  & MuAViC & \underline{8.3} & 20.7 & 24.5 & 21.6 & 18.8 & 18.6 & \underline{13.7} & 19.6 \\
(\textbf{Ours}) Whisper-Flamingo (Large)   & AV & \cmark & MuAViC & \textbf{7.2} & \underline{21.1} & \textbf{25.4} & \underline{22.4} & \underline{19.3} & \underline{19.9} & \textbf{14.7} & \underline{20.5} \\
% \midrule
% \hline
% \rowcolor{Gray}\multicolumn{12}{c}{\textit{Speech-to-Text Translation (Noisy Audio from LRS3)}} \\
% Whisper-Large, Fine-tuned & A & \cmark & LRS3 & 10.8 & 19.7 & 23.2 & 21.8 & 17.9 & 18.2 & 13.7 & 19.1 \\
% \midrule
% (\textbf{Ours}) Whisper-Flamingo (Large) & AV & \cmark & LRS3 & \textbf{5.9} & \textbf{21.4} & \textbf{25.2} & \textbf{22.8} & \textbf{19.6} & \textbf{19.5} & \textbf{14.2} & \textbf{20.5} \\
\hline
\rowcolor{Gray}\multicolumn{12}{c}{\textit{Speech-to-Text Translation (Video only, no audio)}} \\
VSP-LLM~\cite{yeo2024visual} & V & \cmark & - & -  & - &22.7 & 22.3 & 17.9 & 18.7  & - & -\\
\bottomrule
\end{tabular}%
}
\vspace{-0.4cm}
\end{table*}

\subsection{Experimental Setup}
To train our models, we use LRS3~\cite{afouras2018lrs3} -  the largest, publicly-available AVSR dataset in English (En), sourced from TED talks.
We followed AV-HuBERT~\cite{shi2022learning} to create a 433h training set, 1h validation set, and 1h test set.
We also combined the LRS3 training videos with 1,326h of English videos from VoxCeleb2~\cite{chung18b_interspeech} for training.
Transcripts for the VoxCeleb2 videos were obtained from Whisper Large-v2~\cite{vaessen23_interspeech}.
For En-X translation, we use the MuAViC~\cite{anwar23_interspeech} dataset which has translations of LRS3's English text into 6 languages: Greek (El), Spanish (Es), French (Fr), Italian (It), Portuguese (Pt), and Russian (Ru).
% Note that the validation and test sets were manually translated by humans, but most of the training set was machine translated.

We use Whisper Small, Medium, and Large-v2 with 244M, 769M, and 1.55B parameters~\cite{radford2023robust}.
We extract 80-bin log-Mel spectrograms with a stride of 10ms and window size of 25ms from audio sampled at 16kHz.
We extract video features from the AV-HuBERT Large~\cite{shi2022learning} encoder fine-tuned on LRS3 with 325M parameters.
For Whisper Large, the gated cross attention layers add 630M parameters, bringing the total number of parameters to 2.5B (including AV-HuBERT).
We freeze AV-HuBERT but enable dropout and batch normalization updating during Whisper-Flamingo training.
The videos have a frame rate of 25fps and are converted to grayscale.
Dlib~\cite{king2009dlib} is used to extract 96x96 crops centered on the lips which are aligned to a reference mean face~\cite{martinez2020lipreading}.
During training, a random 88x88 crop is used and the video is flipped horizontally with probability 0.5.
For testing, the center 88x88 crop is used.

Table~\ref{tab:hyperparameters} summarizes the hyperparameters for the main experiments.
We used A6000 GPUs with 48GB memory.
Audio/video samples with similar lengths are batched together, and short samples are 0-padded.
AdamW was used as the optimizer ~\cite{loshchilov2018decoupled}.
Following~\cite{radford2023robust}, we used SpecAugment~\cite{park19e_interspeech} (Librispeech-Basic) with Whisper-Large and did not use it with Whisper-Medium.
Training was done with PyTorch~\cite{paszke2019pytorch} and PyTorch Lightning~\cite{Falcon_PyTorch_Lightning_2019}.
We used the SpecAugment and batch sorter implementations from ESPnet~\cite{watanabe18_interspeech}.

We train our models in two conditions: without noise (clean) and with noise (noisy).
For clean training, we do not add any noise to the audio.
For noisy training, we randomly add noise to the audio with a signal-to-noise ratio (SNR) of 0. 
Following prior work~\cite{shi22_interspeech,anwar23_interspeech}, the  ``natural'', ``music'' and ``babble'' noise are sampled from the MUSAN dataset~\cite{snyder2015musan}, and overlapping ``speech'' noise is sampled from LRS3~\cite{afouras2018lrs3}.
To select the best checkpoints, we monitor the highest token prediction accuracy on the clean or noisy validation set every 1k steps.
Following prior work~\cite{anwar23_interspeech}, we use the Fairseq normalizer~\cite{wang-etal-2020-fairseq} to remove punctuation and lower-case text before calculating WER.
For translation, we use SacreBLEU~\cite{post-2018-call} with the default 13-a tokenizer to calculate BLEU~\cite{papineni2002bleu}.

\subsection{Modality Fusion Ablation with Whisper-Medium}
\label{sec:modfusion}
We first compared gated cross attention to early and late fusion using Whisper Medium.
For early fusion, we duplicate AV-HuBERT's 25 Hz video features to temporally align them with Whisper's 50 Hz audio features (after the CNN layers) and use addition to fuse them before Whisper's Transformer encoder.
For late fusion, we use an MLP to fuse the video features with Whisper's audio features after its Transformer encoder.
In both cases, all of Whisper's parameters are fine-tuned.
For audio-only baselines, we use Whisper zero-shot (no fine-tuning) and fine-tuned on LRS3.
We test models in both the clean and noisy conditions with babble-noise injected at 0-SNR.
The results are shown in Table~\ref{tab:fusion}.
Fine-tuning audio-only Whisper decreases the noisy WER of the zero-shot model from 22.2\% to 12.6\%.
We then use the fine-tuned model as initialization to train the models with audio-visual fusion.
Early-fusion obtained a small improvement in both the clean and noisy WERs.
Late-fusion could not fuse the modalities well and performance became worse in both clean and noisy conditions.
Finally, Whisper-Flamingo with gated cross attention obtained the best noisy WER, significantly improving the audio-only Whisper fine-tuned baseline from 12.6\% to  7.0\%, while the clean 
WER was slightly improved from 1.9\% to 1.5\%.
Freezing Whisper helps retain its strong audio skills while new cross attention layers enable it to integrate the visual modality more effectively.

\subsection{Whisper-Flamingo English Speech Recognition}
For our main experiments in Table~\ref{tab:lrs3}, we use Whisper-Large.
We report results using beam search with beam size 1 and 15.
In noisy conditions, we use babble noise at 0-SNR constructed following AV-HuBERT~\cite{shi22_interspeech} by adding audio from 30 speakers from LRS3.
Results for additional noise types and SNR levels are shown in the Appendix, Section~\ref{sec:appendix-noise-lrs3}.

\noindent \textbf{Training on clean audio (without noise)}.
Compared to zero-shot audio-only Whisper-Large, fine-tuning without noise improves the clean ASR WER from 2.1\% to 1.0\% (using LRS3 433h) and 0.68\% (using LRS3+VoxCeleb2 1,759h). 
Our fine-tuned Whisper-Large achieves \textbf{SOTA ASR on LRS3 (0.68\%)} and matches the previous SOTA of 0.7\% from Fast Conformer~\cite{burchi2024multilingual}.
We then use our fine-tuned Whisper models to initialize audio-visual Whisper-Flamingo and achieve 1.1\% / 0.76\% AVSR WER using LRS3 433h / LRS3+VoxCeleb2 1,759h respectively.
Our audio-visual Whisper-Flamingo achieves \textbf{SOTA AVSR on LRS3 (0.76\%)} and matches the current SOTA from Llama-AVSR (0.77\%) using LRS3+VoxCeleb2 and Fast Conformer (0.8\%) using additional data.
Notably, our method only uses 2.5B parameters, while Llama-AVSR uses over 8B parameters.
Also, our method achieves better AVSR performance than LP Conformer (0.9\%) which was trained on 100k videos. 
Fine-tuning Whisper on clean audio does not improve the noisy WER, while training Whisper-Flamingo on clean audio improves the noisy WER.

\noindent \textbf{Training on noisy audio}.
Compared to zero-shot audio-only Whisper-Large, fine-tuning with noise improves the noisy ASR WER from 20.8\% to 11.7\% (using LRS3 433h) and 11.1\% (using LRS3+VoxCeleb2 1,759h). 
The clean ASR WER is slightly worse compared to Whisper fine-tuned without noise, however, the results are close (0.85\% vs 0.68\%).
We then use our fine-tuned Whisper models to initialize audio-visual Whisper-Flamingo and significantly improve the noisy WER to 5.6\% / 5.6\% AVSR WER using LRS3 433h / LRS3+VoxCeleb2 1,759h respectively \textbf{(49.5\% relative noisy WER improvement compared to fine-tuned audio-only Whisper).}
The clean AVSR WER is slightly worse compared to Whisper-Flamingo trained without noise, however, the results are close (0.86\% vs 0.76\%).

\noindent \textbf{Comparing noisy results to SOTA}.
Table~\ref{tab:lrs3} also shows a comparison with prior audio-visual SSL methods and audio-visual fine-tuning methods on LRS3.
Direct comparison in noisy conditions is challenging since different noise datasets were used to generate the babble noise. 
SSL methods AV-HuBERT~\cite{shi22_interspeech}, u-HuBERT~\cite{hsu2022u}, and CMA~\cite{kim2024learning} used LRS3 to generate babble noise, but the noise file they generated was not publicly available.
We followed their procedure to generate the noise, so our noisy conditions are similar but not identical.
Compared with AV-HuBERT, Whisper-Flamingo achieves better clean performance (0.86\% vs 1.4\%) and slightly better noisy results (5.6\% vs 5.8\%), which shows that Whisper-Flamingo is effective at adapting Whisper to the visual features from AV-HuBERT.
Moreover, a major advantage of Whisper-Flamingo over AV-HuBERT is improved translation performance (Section~\ref{sec:translation}).
The best noisy performances are reported by u-HuBERT and CMA; we would like to try them as visual encoders for Whisper-Flamingo, but the weights are not publicly available.
Finally, Whisper-Flamingo outperforms other methods in noise which adapt audio-only models through audio-visual fine-tuning~\cite{may2023audio,simic2023self}, including FAVA-USM~\cite{may2023audio} which was pre-trained on 12M hours of unlabeled audio~\cite{zhang2023google}.
However, the babble noise was generated from different datasets making results not strictly comparable.

\subsection{Whisper-Flamingo En-X Speech Translation}
\label{sec:translation}
For these experiments, we fine-tune Whisper and train Whisper-Flamingo with noise.
We report results with a beam size of 15.

\noindent \textbf{Audio Results}.
In Table~\ref{tab:en-x}, we show the result of fine-tuning audio-only Whisper-Large for En-X translation using the 6 languages in the MuAViC dataset (``Whisper-Large, Fine-tuned'').
Although Whisper was not originally trained for En-X translation, it adapts well to the new task.
Testing with clean audio, we achieve an \textbf{average BLEU score of 22.7}, which outperforms the previous SOTA of 21.9 from Bilingual AV-HuBERT.
Moreover, our model transcribes En audio (WER of 1.5\%) and translates to 6 languages with a single set of parameters, while Bilingual AV-HuBERT fine-tunes separately for each language pair and trains language-specific decoders from scratch.
Our model nearly reaches the text-to-text performance from machine translation models using the ground-truth English text; those models achieve average BLEU scores of 23.1 from a multilingual model and 24.3 from bilingual models.

\noindent \textbf{Audio-Visual Results}.
Once we fine-tune audio-only Whisper for En-X translation, we use it to train Whisper-Flamingo by freezing the weights and adding gated cross attention layers.
Testing with clean audio, Whisper-Flamingo slightly outperforms the audio-only model with an average BLEU score of 22.9 and En WER of 1.3\%.
In noisy conditions, we use multilingual babble noise constructed following MuAViC~\cite{anwar23_interspeech} by adding audio in 9 different languages from 30 speakers.
Note that their noise file was not publicly available, so our noisy conditions are similar but not identical. 
With multilingual babble noise, Whisper-Flamingo significantly outperforms the audio-only Whisper model in average BLEU score (\textbf{20.5 vs 18.6)} and En WER (\textbf{7.2\% vs 13.8\%}).
Compared with the previous SOTA bilingual AV-HuBERT, our audio-only average BLEU is much better (18.6 vs 15.0), but our audio-visual performance is slightly worse (20.5 vs 20.8).
However, our models perform \textbf{both En-X translation and En transcription with a single model}, while their models fine-tune separately for each language pair.
Finally, we show the results using Whisper-Medium and Whisper-Small: Whisper-Flamingo always does better in noise compared to the audio-only baselines, and performance tends to improve as the model size increases.

\noindent \textbf{Comparison with VSP-LLM}.
Finally, we compare with VSP-LLM~\cite{yeo2024visual}, a recent LLM-based approach which uses features from AV-HuBERT as input to an LLM to perform lip-reading and translation using video inputs only (no audio).
Even with noise in the audio, our Whisper-Flamingo (Large) outperforms VSP-LLM on all 4 languages evaluated, which shows the advantage of using both audio and video as inputs instead of just video.
Also, our model only has 2.5B parameters compared with VSP-LLM's 7B parameters.

% V2 - only beam search results
\begin{table}[t]
    \centering
    \caption{Results for English transcription on LRS2. 
    We report results on the original test set (Clean).
    Hours of unlabeled \& labeled audio-visual data used to train each model are shown.
    Note\textsuperscript{\textdagger} that we used AV-HuBERT fine-tuned on LRS3 433h.
    }
    \label{tab:lrs2}
    \vspace{-3mm}
\resizebox{\linewidth}{!}{%
\begin{tabular}{lrrrc}
\toprule
 & \multicolumn{2}{c}{AV Training Hrs} & \multicolumn{2}{c}{Clean WER$\downarrow$} \\
Model & Unlabeled & Labeled & ASR & AVSR \\
\hline
\rowcolor{Gray} \multicolumn{5}{c}{\textit{Audio-Visual methods trained only on transcribed videos}} \\
TM-seq2seq~\cite{afouras2018adeep} & - & 223 & 9.7 & 8.5 \\
CTC/Attention~\cite{petridis2018audio} & - & 380 & 8.3 & 7.0 \\
TDNN~\cite{yu2020audio} & - & 223 & 6.7 & 5.9 \\
CM-seq2seq~\cite{ma2021end} & - & 380 & 3.9 & 3.7 \\
Efficient Conformer~\cite{burchi2023audio} & - & 818 & 2.4 & 2.3 \\
AutoAVSR~\cite{ma2023auto} & - & 3,448 & \textbf{1.5} & \textbf{1.5} \\
\hline
\rowcolor{Gray} \multicolumn{5}{c}{\textit{Audio-Visual SSL methods}} \\
UniAVSR~\cite{pan-etal-2022-leveraging} & - & 380 & 2.7 & 2.6 \\
RAVEn~\cite{haliassos2023jointly} & 1,759 & 223 & 2.5 & - \\
USR~\cite{haliassos2024unified} & 1,759 & 223 & \textbf{1.9} & \textbf{1.9} \\
\hline
\rowcolor{Gray} \multicolumn{5}{c}{\textit{Our audio-only Whisper \textbf{Zero-shot} baselines}} \\
Whisper-Medium,  Beam 1  & - & - & \textbf{5.2} & - \\
Whisper-Large,  Beam 1  & - & - & 5.5 & - \\
\hline
\rowcolor{Gray} \multicolumn{5}{c}{\textit{Our models trained \textbf{with} noise}} \\
Whisper-Medium FT, Beam 1  & - & 223 & \textbf{1.3} & - \\
Whisper-Flamingo, Beam 1  & 1,759 & 223\textdagger & - & \textbf{1.4} \\
\bottomrule
\end{tabular}%
}
\vspace{-0.4cm}
\end{table}

\section{Experiments on LRS2}
We use our framework to conduct experiments on the LRS2 dataset~\cite{afouras2018adeep}.
We followed AutoAVSR~\cite{ma2023auto} to create a 223h training set, 0.6h validation set, and 0.5h test set.
We add noise to the audio while training our models.
We report the ASR and AVSR WER on the LRS2 test set (clean) in Table~\ref{tab:lrs2}.
Comparing zero-shot Whisper Medium and Whisper Large V2, the medium model performs slightly better (5.2\% vs 5.5\%), therefore we use Whisper medium for our experiments.
Our fine-tuned Whisper and Whisper-Flamingo models have 769M and 1.39B parameters respectively.
Compared to zero-shot Whisper, fine-tuned Whisper improves the ASR WER from 5.2\% to \textbf{1.3\%, achieving a new SOTA ASR WER on LRS2.}
Finally, Whisper-Flamingo achieves \textbf{SOTA AVSR WER of 1.4\%}.
Whisper-Flamingo outperforms recent SSL method USR~\cite{haliassos2024unified} (1.9\%) trained with the same hours of video and outperforms AutoAVSR~\cite{ma2023auto} (1.5\%) trained with more hours of video.
Finally, Whisper-Flamingo outperforms Whisper in noisy conditions (shown in the Appendix, Section~\ref{sec:appendix-noise-lrs2}).

\section{Conclusion}
We introduced Whisper-Flamingo, a novel audio-visual model that combines the strengths of AV-HuBERT and Whisper using gated cross attention.
Our audio-visual Whisper-Flamingo significantly outperforms audio-only Whisper in noise.
We showed that Whisper can be fine-tuned for the new task of X-En translation.
Our model performs both En speech recognition and En-X speech translation using one set of parameters while previous methods fine-tune separately on each language.
Our method is a generic way of fusing a visual encoder into the decoder of an ASR model to enable AVSR, and it could work with other models trained on more data in the future.

% \clearpage
% For camera ready
\section{Acknowledgments}
We thank Alex H. Liu, Mohamed Anwar, and the reviewers for helpful discussion.
We thank Videet Mehta for help with the experiments in Table~\ref{tab:noise-full-lrs3}.
This research was supported by the MIT-IBM Watson AI Lab and an NDSEG Fellowship to A.R.

\bibliographystyle{IEEEtran}
\bibliography{mybib}

% \clearpage
\setcounter{table}{0}
\renewcommand{\thetable}{A\arabic{table}}
\section{Appendix}

\subsection{Original Results Table}
\label{sec:appendix-original}
% V2 - only beam search results
\begin{table*}[t]
    \centering
    \caption{This table appeared in the original version of the paper. Results for English transcription on LRS3. 
    We report results on the original test set (Clean) and with babble noise added at 0-SNR (Noisy).
    A= Audio, AV= Audio-Visual.
    Noise dataset= dataset used to make babble noise.
    Hours of unlabeled / labeled data used to train each model are shown.
    Note\textsuperscript{\textdagger} that u-HuBERT~\cite{hsu2022u} and AV-HuBERT~\cite{shi22_interspeech} use a different noise file than us.
    }
    \label{tab:en-original}
    \vspace{-3mm}
\resizebox{0.70\linewidth}{!}{%
\begin{tabular}{lccrrrrrr}
\toprule
 & Test & Noise & \multicolumn{2}{l}{Unlabeled Hrs} & \multicolumn{2}{l}{Labeled Hrs} & \multicolumn{2}{c}{WER$\downarrow$} \\
Model & Modalities & \cellcolor[HTML]{FFFFFF}Dataset & A & AV & A & \multicolumn{1}{l}{AV} & \multicolumn{1}{l}{Clean} & \multicolumn{1}{l}{Noisy} \\
% \midrule
% \hline
% \rowcolor{Gray} \multicolumn{9}{c}{\textit{Supervised Methods}} \\
% Serdyuk et al.~\cite{serdyuk22_interspeech} & AV & NoiseX & - & - & - & 90k & 1.6 & 2.9 \\
% Chang et al.~\cite{chang2022on} & AV & NoiseX & - & - & - & 100k & 0.9 & 1.9 \\
\hline
\rowcolor{Gray} \multicolumn{9}{c}{\textit{Audio-Visual SSL Methods}} \\
AV-BEST-RQ~\cite{may2023audio} & AV & NoiseX & - & \multicolumn{1}{r}{1759} & - & 433 & 2.1 & 6.8 \\
AV2vec~\cite{zhang2023self} & AV & MUSAN & - & \multicolumn{1}{r}{433} & - & 433 & 2.5 & 6.7 \\
AV-HuBERT~\cite{shi22_interspeech} & AV & LRS3{\textdagger} & - & \multicolumn{1}{r}{1759} & - & 433 & 1.4 & 5.8 \\
u-HuBERT~\cite{hsu2022u} & AV & LRS3{\textdagger} & - & \multicolumn{1}{r}{1759} & - & 433 & \textbf{1.3} & \textbf{4.6} \\
% \midrule
\hline
\rowcolor{Gray} \multicolumn{9}{c}{\textit{Audio-Only Pre-train + Audio-Visual Fine-Tune Methods}} \\
Adaptive AV~\cite{simic2023self} & AV & MUSAN & - & \multicolumn{1}{r}{400} & 680k & 30 & 2.3 & 16.3 \\
FAVA~\cite{may2023audio} & AV & NoiseX & \multicolumn{1}{r}{1759} & - & - & 433 & 1.7 & 6.6 \\
FAVA-USM~\cite{may2023audio} & AV & NoiseX & 12M & - & \multicolumn{1}{r}{5000} & 433 & \textbf{1.3} & \textbf{6.2} \\
% \midrule
\hline
\rowcolor{Gray} \multicolumn{9}{c}{\textit{Our Audio-Only Whisper Baselines}} \\
Whisper-Large, Zero-shot (No Fine-Tuning) & A & LRS3 & - & - & 680k & \multicolumn{1}{l}{-} & \textbf{2.1} & 20.8 \\
Whisper-Large, Fine-tuned on LRS3 & A & LRS3 & - & - & 680k & \multicolumn{1}{l}{-} & 2.3 & \textbf{11.7} \\
% \midrule
\hline
\rowcolor{Gray} \multicolumn{9}{c}{\textit{Proposed Audio-Visual Fine-tuning Method}} \\
Whisper-Flamingo (\textbf{Ours}) & AV & LRS3 & - & \multicolumn{1}{r}{1759} & 680k & 433 & \textbf{1.5} & \textbf{5.6}\\
\bottomrule
\end{tabular}%
}
% \vspace{-0.4cm}
\end{table*}

% V2 - only beam search results
\begin{table*}[t]
    \centering
    \setlength{\tabcolsep}{5pt}
    \caption{Results for English transcription on LRS3 433h with different noise types and SNR levels. We use the large versions of our English models fine-tuned with noise and by default report results for beam search decoding with beam size 1. The results for Music and Natural noise from MUSAN are averaged.
    }
    \label{tab:noise-full-lrs3}
    \vspace{-3mm}
\resizebox{\linewidth}{!}{%
\begin{tabular}{llrrrrrrrrrrrrrrrrrrrrrrrrr}
\toprule
Method & \multicolumn{1}{c}{Clean} &  \multicolumn{5}{c}{Babble (LRS3), SNR=} & \multicolumn{5}{c}{Babble (MUSAN), SNR=} & \multicolumn{5}{c}{Babble (MuAViC), SNR=} & \multicolumn{5}{c}{Speech (LRS3), SNR=} & \multicolumn{5}{c}{Music+Natural, SNR=} \\
 & \multicolumn{1}{l}{$\infty$} & -10 & -5 & 0 & 5 & 10 & -10 & -5 & 0 & 5 & 10 & -10 & -5 & 0 & 5 & 10 & -10 & -5 & 0 & 5 & 10 & -10 & -5 & 0 & 5 & 10 \\
\hline
\rowcolor{Gray} \multicolumn{27}{c}{\textit{\textbf{Audio-Only} Testing}} \\
AV-HuBERT~\cite{shi22_interspeech} & 1.6 & \bf{97.5} & \bf{62.3} & 15.7 & 5.1 & 2.6 & - & - & - & - & - & - & - & - & - & - & 81.7 & 56.2 & 37.3 & 19 & 8.3 & 38.7 & 15.1 & 5.7 & 3.1 & 2.3 \\
Whisper, Zero-shot & 2.3 & 101 & 90.2 & 23.3 & 5.5 & 2.9 & 108 & 79.7 & 20.9 & 5.4 & 2.8 & 101 & 89.2 & 28.1 & 6.8 & 2.9 & 101 & 77.4 & 35.6 & 7.6 & \textbf{2.8} & 39.6 & 16.7 & 5.6 & 4.0 & 3.5 \\
$\hookrightarrow$ w/ Beam search 15 & 2.1 & 99.3 & 86.4 & 20.8 & 4.9 & 2.5 & \textbf{98.6} & 73.6 & 19.1 & 5 & 2.6 & \textbf{99.5} & 85.1 & 25.1 & 5.9 & 2.5 & 99.3 & 71.9 & 27.3 & \textbf{5.2} & \textbf{2.8} & 37.0 & 15.1 & 5.2 & 2.8 & 2.3 \\
Whisper, Fine-tuned & \bf{1.0} & 111 & 71.1 & 12.3 & \bf{2.8} & \bf{1.4} & 106 & 57.6 & 10.6 & \bf{2.4} & \bf{1.3} & 107 & 79.1 & 15.1 & \bf{3.5} & \bf{1.4} & 46.1 & 25.9 & \bf{14.9} & 8.4 & 4.6 & 30.8 & \bf{10.0} & \bf{3.0} & \bf{1.5} & \bf{1.2} \\
$\hookrightarrow$ w/ Beam search 15 & 2.7 & 109 & 68.1 & \bf{11.8} & 3.2 & 2.4 & 106 & \bf{55.4} & \bf{10.3} & 3.1 & 2.1 & 110 & \bf{74.8} & \bf{14.3} & 3.9 & 2.3 & \bf{45.8} & \bf{25.2} & 16.0 & 8.7 & 4.6 & \bf{29.4} & \bf{10.0} & 3.5 & 2.5 & 2.3 \\
\hline
\rowcolor{Gray} \multicolumn{27}{c}{\textit{\textbf{Audio-Visual} Testing}} \\
AV-HuBERT~\cite{shi22_interspeech} & 1.4 & 28.4 & 13.4 & 5.0 & 2.6 & 1.9 & - & - & - & - & - & - & - & - & - & - & 11.4 & 4.6 & 2.9 & 1.9 & \bf{1.8} & 9.7 & 4.7 & 2.5 & 1.9 & 1.8 \\
CMA~\cite{kim2024learning} & 1.5 & \bf{25.8} & \bf{11.9} & \bf{4.4} & 2.4 & 1.8 & \bf{22.7} & \bf{9.9} & \bf{4.0} & \bf{2.2} & 1.8 & - & - & - & - & - & \bf{5.4} & \bf{3.2} & \bf{2.5} & \bf{1.8} & \bf{1.8} & \bf{8.7} & \bf{3.7} & 2.4 & 2.0 & 1.7 \\
Whisper-Flamingo & \bf{1.0} & 42.6 & 28.0 & 6.3 & \bf{2.0} & \bf{1.4} & 44.2 & 22.6 & 5.8 & 2.1 & \bf{1.3} & 49.3 & 28.4 & 7.2 & \bf{2.5} & \bf{1.5} & 23.8 & 13.4 & 8.1 & 4.7 & \bf{1.8} & 14.0 & 4.8 & \bf{2.0} & \bf{1.4} & \bf{1.2} \\
$\hookrightarrow$ w/ Beam search 15 & 1.5 & 37.7 & 23.5 & 5.6 & 2.1 & 1.6 & 37.5 & 19.8 & 5.0 & 2.1 & 1.4 & \bf{40.2} & \bf{25.2} & \bf{6.6} & \bf{2.5} & \bf{1.5} & 21.2 & 13.0 & 6.9 & 3.4 & \bf{1.8} & 11.7 & 4.9 & 2.3 & 1.7 & 1.6 \\
\bottomrule
\end{tabular}%
}
\end{table*}

% V2 - only beam search results
\begin{table*}[!]
    \centering
    \setlength{\tabcolsep}{5pt}
    \caption{Results for English transcription on LRS2 223h with different noise types and SNR levels. We use the medium versions of our English models fine-tuned with noise and by default report results for beam search decoding with beam size 1. The results for Music and Natural noise from MUSAN are averaged.
    }
    \label{tab:noise-full-lrs2}
    \vspace{-3mm}
\resizebox{\linewidth}{!}{%
\begin{tabular}{llrrrrrrrrrrrrrrrrrrrrrrrrr}
\toprule
Method & \multicolumn{1}{c}{Clean} &  \multicolumn{5}{c}{Babble (LRS3), SNR=} & \multicolumn{5}{c}{Babble (MUSAN), SNR=} & \multicolumn{5}{c}{Babble (MuAViC), SNR=} & \multicolumn{5}{c}{Speech (LRS3), SNR=} & \multicolumn{5}{c}{Music+Natural, SNR=} \\
 & \multicolumn{1}{l}{$\infty$} & -10 & -5 & 0 & 5 & 10 & -10 & -5 & 0 & 5 & 10 & -10 & -5 & 0 & 5 & 10 & -10 & -5 & 0 & 5 & 10 & -10 & -5 & 0 & 5 & 10 \\
\hline
\rowcolor{Gray} \multicolumn{27}{c}{\textit{\textbf{Audio-Only} Testing}} \\
Whisper, Zero-shot & 5.2 & 103 & 98.9 & 31.7 & 10.1 & 5.5 & 117 & 103.5 & 42.3 & 11.3 & 6.2 & 111 & 88.3 & 34.8 & 9.9 & 5.9 & 96.1 & 72.2 & 30.9 & 9.9 & 5.1 & 56.1 & 27.6 & 10.9 & 6.4 & 5.5 \\
$\hookrightarrow$ w/ Beam search 15 & 3.9 & \bf{98.7} & 83.4 & 29.2 & 8.2 & 4.7 & \bf{99.8} & 91.5 & 37.7 & 10.1 & 4.8 & \bf{100} & 80.7 & 29.2 & 8.7 & 4.8 & 98 & 73.6 & 27.4 & 7.7 & 4.7 & 45.9 & 23.3 & 9.4 & 5.2 & 4.3 \\
Whisper, Fine-tuned & \bf{1.3} & 136 & 73.5 & 15.5 & 4.5 & \bf{2.2} & 144 & 94.1 & 20.6 & \bf{4.4} & \bf{2.0} & 134 & 68.2 & 13.8 & \bf{4.1} & \bf{2.2} & 34 & 15.5 & 6.3 & \bf{2.9} & \bf{2.0} & 36.3 & 14.1 & \bf{4.9} & \bf{2.1} & \bf{1.7} \\
$\hookrightarrow$ w/ Beam search 15 & 1.7 & 114 & \bf{70.4} & \bf{14.5} & \bf{4.4} & 2.3 & 125 & \bf{89.2} & \bf{18.8} & 4.5 & 2.3 & 115 & \bf{62.1} & \bf{12.5} & \bf{4.1} & 2.5 & \bf{31.7} & \bf{13.7} & \bf{6.1} & 3.2 & 2.3 & \bf{35.8} & \bf{13.9} & 5.1 & 2.8 & 2.2 \\
\hline
\rowcolor{Gray} \multicolumn{27}{c}{\textit{\textbf{Audio-Visual} Testing}} \\
Whisper-Flamingo & \bf{1.4} & \bf{102} & 65.5 & 12.9 & \bf{4.1} & \bf{2.0} & 123 & 83.1 & 16.4 & \bf{3.9} & \bf{2.1} & \bf{103} & 55.8 & \bf{11.9} & \bf{3.5} & \bf{1.9} & 29.5 & 13.1 & \bf{5.5} & \bf{2.8} & \bf{2.0} & 31.1 & 11.5 & \bf{4.1} & \bf{2.1} & \bf{1.6} \\
$\hookrightarrow$ w/ Beam search 15 & 2.1 & 103 & \bf{59.9} & \bf{12.2} & 4.3 & 2.4 & \bf{118} & \bf{78.2} & \bf{15.1} & 4.1 & 2.5 & 106 & \bf{51.8} & 12.0 & 3.6 & 2.3 & \bf{27.8} & \bf{12.6} & 6.1 & 3.8 & 2.6 & \bf{30.1} & \bf{10.9} & 4.2 & 2.7 & 2.2 \\
\bottomrule
\end{tabular}%
}
\end{table*}

Table~\ref{tab:en-original} shows the results reported in the original paper (V1 on ArXiv). 
For these results, beam search with a beam size of 15 was used.
Note that the current results for Whisper-Flamingo in Table~\ref{tab:lrs3} are a superset of those reported in Table~\ref{tab:en-original} (the original results remained the same and we added more results in the current main table).

% V2 - only beam search results
\begin{table}[t]
    \centering
    \caption{Analysis of position to insert gated cross attention layers into Whisper. 
    All models are trained on LRS3 433h and evaluated with babble noise added at 0-SNR. 
    % When inserting the layers into Whisper encoder blocks, we simultaneously insert the layers in the beginning of the Whisper decoder blocks.
    }
    \label{tab:x_attn_analysis}
    \vspace{-3mm}
\resizebox{0.40\linewidth}{!}{%
\begin{tabular}{lr}
\toprule
Position & Noisy \\
 & WER \\
\hline
\rowcolor{Gray} \multicolumn{2}{c}{\textit{Whisper decoder blocks}} \\
Beginning & \textbf{5.6} \\
After self-attention & 6.1 \\
After cross-attention & 6.3 \\
After MLP & 5.7 \\
\hline
\rowcolor{Gray} \multicolumn{2}{c}{\textit{Whisper encoder blocks}} \\
Beginning & 6.1 \\
After self-attention & 6.1 \\
After MLP & \textbf{5.8} \\
\bottomrule
\end{tabular}%
}
\end{table}

\subsection{LRS3: Testing Different Noise Types and Noise Levels}
\label{sec:appendix-noise-lrs3}
Table~\ref{tab:noise-full-lrs3} shows the results on LRS3 433h with different noise types and SNR levels $\{-10,-5,0,5,10,\infty \}$. 
The noise setups follow AV-HuBERT~\cite{shi22_interspeech} and CMA~\cite{kim2024learning}, and we also test on multilingual babble noise that we made using audio from MuAViC.
However, some of the differences in the results across methods might be due to different random seeds since the noise files are sampled from a list of candidates (except for babble noise from LRS3 and babble noise from MuAViC which only use 1 noise file for testing).
We first compare models testing with audio only and then compare models testing with both audio and visual modalities.
We use the Large versions of our English models fine-tuned with noise and report decoding results with beam size 1 and 15.
We make the following observations:
\begin{itemize}
    \item Audio-visual Whisper-Flamingo significantly outperforms audio-only Whisper (fine-tuned) with both beam size 1 and 15.
    \item Comparing our results with beam size 1 and 15, beam size 15 tends to outperform beam size 1 for SNRs $\leq 0$, but beam size 1 performs better for SNRs $> 0$.
    \item For audio-only testing, our fine-tuned Whisper outperforms zero-shot Whisper for almost all noise types and levels. Whisper (fine-tuned) tends to outperform AV-HuBERT~\cite{shi22_interspeech}.
    \item For audio-visual testing, AV-HuBERT and CMA~\cite{kim2024learning} tend to outperform Whisper-Flamingo for SNRs $\leq 0$, while Whisper-Flamingo tends to outperform them for SNRs $> 0$.
    \item The ordering of performance on the three babble noise types from worst to best are: (1) MuAViC (2) LRS3 (3) MUSAN. 
    We believe that MuAViC babble noise is more difficult since it is multilingual.
    \item Comparing the difficulty of babble noise with the other types, speech from LRS3 is also difficult at 0 SNR, but easier than babble at SNRs $\leq 0$.
    Also, music and natural noise are easier than babble noise.
\end{itemize}
Overall, the results confirm Whisper-Flamingo's noise robustness using audio and visual modalities. 
Future work can focus on enhancing the performance in low SNR conditions and comparing more comprehensively with prior work~\cite{afouras2018adeep,xu2020discriminative,hu2023cross,hu2023hearing,hu2023mir,chen2023leveraging}.

\subsection{LRS2: Testing Different Noise Types and Noise Levels}
\label{sec:appendix-noise-lrs2}
Table~\ref{tab:noise-full-lrs2} shows the results on LRS2 223h with different noise types and SNR levels $\{-10,-5,0,5,10,\infty \}$. 
The noise setups follow AV-HuBERT~\cite{shi22_interspeech}, and we also test on multilingual babble noise that we made using audio from MuAViC.
We compare Whisper Medium zero-shot, Whisper Medium fine-tuned, and Whisper-Flamingo.
We report decoding results with beam size 1 and 15.
In general, the observations follow those in Section~\ref{sec:appendix-noise-lrs3}.
Most importantly, fine-tuned Whisper outperforms zero-shot Whisper, and Whisper-Flamingo outperforms fine-tuned Whisper in most noise types and levels.
However, Whisper-Flamingo's improvements over fine-tuned Whisper are not as substantial compared to the improvements achieved on the LRS3 dataset (Table~\ref{tab:noise-full-lrs3}).
A potential reason could be that we use AV-HuBERT trained on LRS3 and VoxCeleb2, and we do not update its weights while training Whisper-Flamingo.
Future work should direct more attention to improving the noise robustness of Whisper-Flamingo on LRS2.

\subsection{Analysis of Gated Cross Attention Position}
\label{sec:appendix-x-attn}
Table~\ref{tab:x_attn_analysis} shows the analysis of the position to insert the gated cross attention layers into Whisper.
Refer to Figure~\ref{fig:model_fig} for the ordering of layers in the transformer decoder and encoder blocks.
We first tried all possible positions in Whisper's decoder blocks and found the beginning of each block to work the best (5.6\% noisy AVSR WER), although after the MLP in each block also worked well (5.7\% noisy AVSR WER).
We then tried to insert the layers in both Whisper's decoder and encoder blocks, but this did not provide any additional gains.

\subsection{Analysis of Training With and Without Noise}
\label{sec:appendix-noise}
% V2 - only beam search results
\begin{table*}[t]
    \centering
    \caption{Full analysis of clean vs noisy training for Whisper-Large and Whisper-Flamingo. 
    All models are trained on LRS3 433h.
    We report results on the original test set (Clean) and with babble noise added at 0-SNR (Noisy).
    Init.=Initialized.
    }
    \label{tab:en_analysis}
    \vspace{-3mm}
\resizebox{0.6\linewidth}{!}{%
\begin{tabular}{lrrrr}
\toprule
 & \multicolumn{2}{c}{ASR WER$\downarrow$} & \multicolumn{2}{c}{AVSR WER$\downarrow$} \\
Model & Clean & Noisy & Clean & Noisy \\
\hline
\rowcolor{Gray} \multicolumn{5}{c}{\textit{Our audio-only Whisper baselines}} \\
Whisper, Zero-shot, Beam 1 & 2.3 & 23.3 & - & - \\
$\hookrightarrow$ w/ Beam search 15 & 2.1 & 20.8 & - & - \\
Whisper, Fine-tuned \textbf{w/o Noise}, Beam 1 & \textbf{1.0} & 21.6 & - & - \\
$\hookrightarrow$ w/ Beam search 15 & 1.3 & 23.1 & - & - \\
Whisper, Fine-tuned \textbf{w/ Noise}, Beam 1 & 1.1 & 12.4 & - & - \\
$\hookrightarrow$ w/ Beam search 15 & 2.3 & \textbf{11.7} & - & - \\
\hline
\rowcolor{Gray} \multicolumn{5}{c}{\textit{Whisper-Flamingo trained \textbf{without} noise}} \\
Init. from ``Whisper, Zero-shot'', Beam 1 & - & - & 13.5 & 48.6 \\
$\hookrightarrow$ w/ Beam search 15 & - & - & 15.3 & 31.9 \\
Init. from ``Whisper, Fine-tuned \textbf{w/o Noise}'', Beam 1 & - & - & 1.1 & 13.0 \\
$\hookrightarrow$ w/ Beam search 15 & - & - & 1.1 & 12.8 \\
Init. from ``Whisper, Fine-tuned \textbf{w/ Noise}'', Beam 1 & - & - & \textbf{1.0} & 8.1 \\
$\hookrightarrow$ w/ Beam search 15& - & - & 1.2 & \textbf{7.3} \\
\hline
\rowcolor{Gray} \multicolumn{5}{c}{\textit{Whisper-Flamingo trained \textbf{with} noise}} \\
Init. from ``Whisper, Zero-shot'', Beam 1 & - & - & 12.9 & 32.8 \\
$\hookrightarrow$ w/ Beam search 15 & - & - & 13.8 & 29.0 \\
Init. from ``Whisper, Fine-tuned \textbf{w/o Noise}'', Beam 1 & - & - & 1.1 & 9.6 \\
$\hookrightarrow$ w/ Beam search 15 & - & - & 1.3 & 8.5 \\
Init. from ``Whisper, Fine-tuned \textbf{w/ Noise}'', Beam 1 & - & - & \textbf{1.0} & 7.2 \\
$\hookrightarrow$ w/ Beam search 15& - & - & 1.5 & \textbf{5.6} \\
\bottomrule
\end{tabular}%
}
% \vspace{-0.4cm}
\end{table*}

Table~\ref{tab:en_analysis} shows the full analysis of fine-tuning Whisper-Large and training Whisper-Flamingo with and without noise on LRS3 433h.
The results are the same as in Table~\ref{tab:lrs3} with additional results for audio-visual Whisper-Flamingo.
We first present the audio-only Whisper baselines: zero-shot (no fine-tuning), fine-tuned without noise, and fine-tuned with noise.
We report results using beam search with a beam size of 1 (greedy) and 15.
For Whisper-Flamingo audio-visual training, we initialize the model using each of our three Whisper models and add the new gated cross attention layers.
We then train Whisper-Flamingo with noise or without noise.
The takeaways are as follows.
First, initializing Whisper-Flamingo from Whisper zero-shot does not work well. 
The performance is poor even testing on clean audio.
We believe that Whisper-Flamingo must solve three problems to work effectively: 1. it needs to handle noisy audio, 2. it needs to adapt to the specific domain of interest (Ted talks), 3. it needs to integrate a new modality (visual features).
Solving all three of these tasks might be too difficult for Whisper-Flamingo without fine-tuning Whisper.
Fine-tuning Whisper solves the first two problems and enables Whisper-Flamingo to focus on adapting to visual features.
Next, for Whisper-Flamingo trained on clean audio (without noise), initializing from either Whisper fine-tuned without noise or Whisper fine-tuned with noise both achieve good clean AVSR WER (1.1\% and 1.0\% WER respectively).
Interestingly, Whisper-Flamingo always achieves better noisy WER than the corresponding fine-tuned audio-only Whisper, which shows the advantage of using both audio and video modalities.
Finally, for Whisper-Flamingo trained on noisy audio, initializing from Whisper fine-tuned with noise performs better than initializing from Whisper fine-tuned without noise.
The former achieves 5.6\% AVSR WER while the latter achieves 8.5\% AVSR WER.
Overall, Whisper-Flamingo trained with noise achieves the best noisy WER (5.6\%) and matches the clean WER of our best fine-tuned Whisper model (1.0\% WER).

\end{document}